\documentclass{article}      

\title{An Example Document}  
\author{Leslie Lamport}      
\date{January 21, 1994}      

\begin{document}

\title{Dynamical study of spinodal \\ decomposition in heavy ion collisions}

\author{A. Barra\~n\'on${}^1$ and J. A. L\'opez${}^2$}

\author{   A.  Barra\~n\'on  
\footnote{ Universidad Aut\'onoma Metropolitana. Unidad Azcapotzalco.
Av. San Pablo 124, Col. Reynosa-Tamaulipas, Mexico City. email: bca@correo.azc.uam.mx } ;
 J. A.  L\'opez
\footnote{Dept. of Physics, The University of Texas at El Paso. El Paso, TX, 79968  }   } 

\date{Nov, $2^{nd}$,  2004}

\maketitle

\abstract

Nuclei undergo a phase transition in nuclear reactions
according to a caloric curve determined by the amount of entropy.
Here, the generation of entropy is studied in relation to the size
of the nuclear system.

\section{Introduction}
Collisions at hundreds of $MeV$s break the nuclei into several
fragments\cite{mastinu96}. This fragmentation led to the
determination\cite{pocho97,hirschegg99} of the relation between the
system's temperature and its excitation energy, {\it i.e.} the
caloric curve, $CC$. A striking feature of the $CC$ is the
dependence of the transition temperature, {\it i.e.} the ``plateau''
temperature $T_p$, with the mass of the breaking system.  Recent
observations\cite{natowitz2002} show that lighter systems, {\it
i.e.} with masses between $30$ to $60$ nucleons have a $T_p \approx
9 \ Mev$, while systems with masses between $180$ to $240$ yield a
$T_p \approx 6 \ MeV$. This variation of $T_p$ has been linked to
the entropy generated during the reaction\cite{barra2004,brasil}.
Here these collisions are used to find the relationship between the
entropy generated in the reaction and the size of the fragmenting
system.

\section{Caloric curve and entropy}\label{tp}
Ions fuse in collisions and reach some maximum density and
temperature, to then expand cooling and reducing the
density\cite{bertsch,lopezsiemens}. In the density-temperature
plane\cite{lopezdorso}, this corresponds to a displacement from
normal saturation density and zero temperature, $(n_o\approx 0.15 \
fm^{-3}, \ T=0)$, to some higher values of $(n,T)$, to then approach
$n\rightarrow 0$ and $T\rightarrow 0$ asymptotically.

These reactions include phase changes and non-equilibrium dynamics,
and are best studied with molecular dynamics. The model we use is
``$LATINO$''\cite{latino}, combined with the fragment-recognition
code $ECRA$\cite{ecra}; these, without adjustable parameters, have
described reactions\cite{chernolaval}, phase
transitions\cite{oax2001}, criticality\cite{HIP2003}, and the
$CC$\cite{barra2004,brasil}.

The right panel of figure 1 shows trajectories in the $n-T$
plane of collisions simulated with ``$Latino$'' and corresponding to
central collisions of $Ni+Ni$ at energies between $800$ to $1600 \
MeV$. As during the reaction, nucleons evaporate leaving a smaller
system, the trajectories follow different paths on the plane
according to the residual size of the nuclear system. The plot shows
the cases of systems with $49$, $75$, and $91$ nucleons.

As noted before\cite{barra2004,brasil}, these trajectories are
determined by the entropy generated in the beginning of the
reaction. These reactions can help to relate the caloric curves to
system size, and to the amount of entropy generated.

\subsection{Mass dependence of the caloric curve}
Central collisions were performed for $Ni+Ag$ at different beam
energies. During the collision, $n$ and $T$ of the $50 \%$ most
central particles were determined as a function of time. At the same
time, $ECRA$ was used to identify the fragment source size,
fragmentation times, fragment stability, and other quantities of
interest. This information was then used to obtain the caloric curve
as a function of the source size.

The left panel of figure 2 shows caloric curves obtained by
$Latino$ in central collisions of $Ni+Ag$ at energies between $1000$
to $3800 \ MeV$. The three groups correspond to different masses:
circles for collisions with residual sources of $40$ to $59$
nucleons, rectangles for $60$ to $79$, and crosses for $80$ to $100$
nucleons. The symbol sizes denote the standard deviations of $T$ and
excitation energy of each energy bin. The lines, average values of
$T$ of the last few points of each mass range, clearly show the
inverse relationship between the transition temperature, $T_p$, and
source size, as observed experimentally\cite{natowitz2002}.

\subsection{Entropy dependence of the caloric curve}
The collision takes a $T=0$ quantum-system to a hot and dense
``quasi-classical'' state through a non-equilibrium path, which then
expands and decomposes spinodaly.  In terms of entropy, the system
goes from an initial $S=0$ state to a high-entropy state, to then
coast into an isentropic expansion until a violent spinodal
decomposition increases entropy again.

$Latino+ECRA$ allow the evaluation of the entropy during the
reaction.  Neglecting quantum effects, the entropy per particle $S$
can be quantified in units of Boltzmann's constant as for a
classical gas\cite{huang} through  $S = \log
\left[{{1}\over{n}}\left({{3T}\over{2}}\right)^{3/2}\right]+ S_o$,
where $S_o$ depends on the nucleon mass\cite{barra2004}.

Figure 2 shows the entropy of central collisions of $Ni+Ni$
at energies between $600$ to $2000 \ MeV$ plotted against the size
of the residual fragmenting source. The bottom curve corresponds to
the entropy achieved at maximum heating and compression, {\it i.e.}
the value at which the system expands until it decomposes; the one
responsible for determining $T_p$.  The top curve, on the other
hand, is the asymptotic value of $S$, and it includes all other
increases of entropy that took during the phase change and
expansion.  Both of these curves show that lighter systems are more
apt to generate entropy, and that the phase change and final
expansion succeed in increasing the initial entropy.

\section{Conclusions}
These results confirm previous results\cite{barra2004,brasil}, the
initial stage of the reaction reaches a value of $S$ which defines
the trajectory of the compound nucleus into a spinodal
decomposition. The transition temperature $T_p$ is thus defined by
the intersection of the isentrope and the spinodal.  The present
study further confirms that the amount of entropy generated
initially in the reaction varies inversely with the size of the
fragmenting system. This inverse relationship is also maintained
after the phase change. Unfortunately, as concluded in previous
studies, the observed relationship between $S$ and source size was
obtained with $LATINO$, and does not explain what causes lighter
systems to generate more entropy than heavier ones in heavy ion
collisions. We propose to continue with this study in the near
future.

\end{document}